\documentclass[a4paper]{jpconf}
\usepackage{color,graphicx,epsfig}
\usepackage{amsmath,amssymb,amsfonts,dsfont,xcolor}

\usepackage{biblatex}
\newcommand{\btosll}{\ensuremath{b\to s\ell\ell} }
\newcommand{\btosmumu}{\ensuremath{b\to s\mu^+\mu^-} }
\newcommand{\stodnunu}{\ensuremath{s\rightarrow d\nu\bar\nu} }
\newcommand{\btosnunu}{\ensuremath{b\rightarrow s\nu\bar\nu} }

\usepackage{hyperref}
\begin{document}
\title{Connecting \btosll with \btosnunu and \stodnunu}

\author{Mart\'in Novoa-Brunet}

\address{Istituto Nazionale di Fisica Nucleare, Sezione di Bari, Via Orabona 4, 70126 Bari, Italy}

\ead{martin.novoa@ba.infn.it}

\begin{abstract}
We discuss the consequences of deviations from the Standard Model observed in \btosmumu transitions
for flavour-changing neutral-current processes involving down-type quarks and neutrinos. We work within an effective field theory approach respecting the SM gauge
symmetry, including right-handed currents, a flavour structure based on approximate U(2) symmetry,
and assuming only SM-like light neutrinos. We discuss correlations among $B\to h_s\nu\bar\nu$ ($h_s=K,K^*,X_s$), $K^+\to\pi^+\nu\bar\nu$ and $K_L\to\pi^0\nu\bar\nu$
branching ratios in the case of linear Minimal Flavour Violation and in a more general framework.\\[5pt]
{\bfseries Report Number.}\quad\rm\ignorespaces  BARI-TH/22-739
\end{abstract}

\section{Introduction}\label{sec:introduction}
In recent years, several deviations from the SM predictions in semi-leptonic b-quark decays~\cite{Bifani:2018zmi} data have appeared hinting towards New Physics (NP) and Lepton Flavour Universality Violation (LFUV). More specifically, consistent deviations have been observed for the $\btosll$ transition on Branching fractions and Angular observables for the $B\to K^{(*)} \mu^+\mu^-$ and $B_s\to \phi \mu^+\mu^-$ modes. Furthermore, the deviations in the so-called LFU Ratios ($R_K^{(*)}$)  are a strong hint for Lepton Flavour Universality Violation (LFUV).

These deviations can be interpreted model-independently in terms of specific contributions to short-distance Wilson coefficients $\mathcal{C}_i$ of the weak effective Hamiltonian \cite{Buchalla:1995vs,Buras:1998raa}. The most relevant Wilson coefficients to explain LFUV in \btosll are the ones corresponding to the Vector/Axial current operators $\mathcal{O}_{9,10}^{(\prime)\ell}$. 
The global fits to $b\to s\ell\ell$ data \cite{Alguero:2021anc} show that these deviations exhibit a consistent pattern favouring significant non-standard effects in muonic final states ($\mathcal{C}_{9,10}^\mu$). While a smaller effect in electrons is not excluded it is not required to fit data ($\mathcal{C}_{9,10}^e$) and $b\to s\tau^+\tau^-$ transitions are at present only poorly constrained but could appear through LFU contributions equal for all lepton flavours ($\mathcal{C}_{9}^\ell$) \cite{Capdevila:2017iqn}. The most prominent 1D/2D scenarios which can explain the \btosmumu anomalies contain either a pure $\mathcal{C}^{\mu,{\rm NP}}_9$ (left-handed quark current) contribution or a combination of the former with either $\mathcal{C}^{\mu,{\rm NP}}_{10}$ (left-handed quark current) or $\mathcal{C}^{\mu,{\rm NP}}_{9'}$ (right-handed quark current).

Since SM neutrinos reside in the same leptonic weak doublets as the left-handed charged leptons, decay modes with neutrinos in the final state are a natural place where related NP effects can show, most evidently $b\to s \nu\bar\nu$ decays or also other Flavour-Changing Neutral Currents (FCNC), namely $s\to d \nu\bar \nu$ decays, since they can be related to $\btosll$ through approximate flavour symmetries.
Furthermore, the $B\to h_s \nu \bar \nu$ ($h_s=K,K^*,X_s$) and $K\to \pi \nu \bar \nu$ decays are known for their NP sensitivity~\cite{Altmannshofer:2009ma}.

 The main goal of our approach is to determine the impact of the current \btosll results on future measurements of $B \to h_s \nu \bar \nu$ and $K\to \pi \nu \bar \nu$ in a general effective theory framework and to illustrate the potential correlations among these measurements. This contribution is based on Ref.~\cite{Descotes-Genon:2020buf}.

\section{NP in semileptonic FCNC decays}

Possible heavy NP contributions are written in terms of $SU(2)_L$ gauge invariant operators~\cite{Buttazzo:2017ixm}
\begin{equation}
  \begin{split}
    & \mathcal L_{\rm eff.}   = \mathcal L_{\rm SM}  - \frac{1}{v^2} \lambda^q_{ij} \lambda^\ell_{\alpha\beta} \left[ C_T \left( \bar Q_L^i \gamma_\mu \sigma^a Q_L^i \right) \left( \bar L_L^\alpha \gamma^\mu \sigma^a L_L^\beta \right) +  C_S \left( \bar Q_L^i \gamma_\mu  Q_L^i \right) \left( \bar L_L^\alpha \gamma^\mu  L_L^\beta \right)   \right.\\
    &\left.  + C'_{RL} \left( \bar d_R^i \gamma_\mu d_R^i \right) \left( \bar L_L^\alpha \gamma^\mu  L_L^\beta \right) + C'_{LR} \left( \bar Q_L^i \gamma_\mu  Q_L^i \right) \left( \bar \ell_R^\alpha \gamma^\mu  \ell_R^\beta \right)  + C'_{RR} \left( \bar d_R^i \gamma_\mu d_R^i \right) \left( \bar \ell_R^\alpha \gamma^\mu  \ell_R^\beta \right)
    \right]\,, \label{eq:ops}
  \end{split}
\end{equation}
where $Q_L^i = (V^{\rm CKM *}_{ji}  u_L^j, d_L^i)^T$
and $L_L^\alpha = (U^{\rm PMNS}_{\alpha\beta}\nu_L^\beta,\ell^{\alpha}_L)^T$.

The presence of operators with lepton doublets in Eq.~(\ref{eq:ops}) is the basis for the connection between flavour-changing neutral currents involving charged and neutral leptons that we explore in the following.
We assume that the same flavour structure encoded in $\lambda^q_{ij}$ (quark sector) and $\lambda^\ell_{\alpha\beta}$ (lepton sector) holds for all operators.

Regarding the quark sector, we classify the NP flavour structures in terms of an approximate  $U(2)_{q=Q, D}$ flavour symmetry acting directly on the quark fields, under which two generations of quarks form doublets, while the third generation is invariant. We will focus on down-type quarks. One can write
${\bf q}\equiv (q_L^1, q_L^2) \sim ({\bf 2},{\bf 1})$, ${\bf d} \equiv (d_R^1,d_R^2) \sim ({\bf 1},{\bf 2})$ while $ d_R^3,q_L^3 \sim ({\bf 1},{\bf 1})$\,.
In the exact $U(2)_{q}$ limit only $\lambda^q_{33}$ and $\lambda^{q}_{11}=\lambda^q_{22} $ in Eq.~\eqref{eq:ops} are non-vanishing.
To avoid excessive effects in neutral kaon oscillation observables, we thus furthermore impose the {\it leading} NP $U(2)_q$ breaking to be aligned with the SM Yukawas, yielding a General Minimal Flavour Violating (GMFV)~\cite{Kagan:2009bn} structure with the singlet field defined as
$d _{L}^3 =  b_L+ \theta_q e^{i\phi_q} \left( V_{td} d_L   + V_{ts}  s_L \right)$, whereas $d^1_L=d_L$, $d^2_L=s_L$ (and similarly for $q^3_L$), where $\theta_q$ and $\phi_q$ are fixed but otherwise arbitrary numbers.
The linear MFV limit~\cite{Kagan:2009bn} is recovered by taking $\theta_q = 1$ and $\phi_q=0$ (taking $V_{tb}=1$).

For the lepton sector we assume an approximate $U(1)^3_\ell$ symmetry (broken only by the neutrino masses) yielding $\lambda_{i\neq j}^\ell \simeq 0$\, as required by stringent limits on lepton flavour violation~\cite{Davidson:2006bd,Glashow:2014iga,Alonso:2015sja}. We consider here only (SM-like) left-handed neutrinos.
 Since neutrino flavours are virtually impossible to tag in rare meson decay experiments, we need to assume specific ratios of $U(1)_\ell^3$ charges ($\lambda^{\ell}$) in order to correlate FCNC processes involving charged leptons and neutrinos. We consider three well known scenarios: scenario 1) $\lambda^{\ell}_{ee} =\lambda^{\ell}_{\tau\tau} = 0$ NP effects only in muonic final states, scenario 2) $\lambda^{\ell}_{\mu\mu} = -\lambda^{\ell}_{\tau\tau}$ and $\lambda^{\ell}_{ee} =0$ well suited for UV-complete model building~\cite{Altmannshofer:2014cfa,Crivellin:2016ejn} and scenario 3) $\lambda^{\ell}_{\alpha\alpha} \propto m_\alpha $ with $\alpha=e,\, \mu,\,\tau$  hierarchical scenario~\cite{Redi:2011zi,Niehoff:2015bfa}.

\section{Charged vs Neutral Lepton modes}

The Lagrangian in Eq.~\eqref{eq:ops}, which extends the one considered in Ref.~\cite{Buttazzo:2017ixm} to include right-handed fields, is chosen to reproduce the results of the global fits to $b\to s\ell\ell$ data discussed in Sec.~\ref{sec:introduction}. Scalar and tensor operators are excluded since they do not provide a good fit for $b\to s\ell\ell$ data~\cite{Alguero:2021anc}. At leading order in $U(2)_q$ breaking the relevant Wilson coefficients for each mode are given by\footnote{The expressions of the branching fractions and the full expression of the weak effective Hamiltonian Wilson coefficients as a function of the coefficients of Eq.~\eqref{eq:ops} can be found in Ref.~\cite{Descotes-Genon:2020buf}.}:
\\
\btosll:
\begin{equation}
 \begin{split}
   C^{\mu,{\rm NP}}_{9}   & \propto  \lambda_{\mu\mu}^\ell \lambda_{33}^q \left({(C_T +C_S)} +C_{LR}' \right)  +\mathcal O (\lambda_{23}^q)  \\
   C^{\mu,{\rm NP}}_{10}  & \propto  \lambda_{\mu\mu}^\ell \lambda_{33}^q\left(- { (C_T +C_S)} +C_{LR}' \right) + \mathcal O (\lambda_{23}^q) \\
   C^{\mu,{\rm NP}}_{9'}  & \propto  \lambda_{\mu\mu}^\ell { \lambda_{23}^q } \left( C_{RR}'+C_{RL}'   \right)                               \\
   C^{\mu,{\rm NP}}_{10'} & \propto  \lambda_{\mu\mu}^\ell { \lambda_{23}^q }   \left(C_{RR}'  - C_{RL}' \right)
   \label{eq:C9fromGMFV}
 \end{split}
 \end{equation}
\btosnunu:
\begin{equation}
 \begin{split}
   C^{\nu_\alpha, {\rm NP}}_L & \propto  \lambda_{\alpha\alpha}^\ell  \lambda_{33}^q   { (C_T -C_S)}  +\mathcal O (\lambda_{23}^q) \\
   C^{\nu_\alpha, {\rm NP}}_R & \propto  \lambda_{\alpha\alpha}^\ell  { \lambda_{23}^q }  C'_{RL}
 \end{split}
 \end{equation}
\stodnunu:
  \begin{align}
   C^{\nu_\alpha,{\rm NP}}_{sd} & \propto  \lambda_{\alpha\alpha}^\ell \lambda_{33}^q   { (C_T -C_S)}  +\mathcal O (\lambda_{23}^q)  +\mathcal O (\lambda_{13}^q)
  \end{align}

In (G)MFV,  right-handed FCNCs among down-type quarks are suppressed so that we may set $C'_{RL}=C'_{RR}=0$ then.
Departures from the (G)MFV limit may manifest through additional
explicit $U(2)_q$ breaking effects appearing as $\lambda^q_{i\neq j} \neq 0$ as seen above. From Eq.~\eqref{eq:C9fromGMFV} we can see that, the $C^{\mu,{\rm NP}}_{9}$ and the $(C^{\mu,{\rm NP}}_{9}, C^{\mu,{\rm NP}}_{10})$ scenarios coming from \btosll global fits can be accommodated in GMFV while the $(C^{\mu,{\rm NP}}_{9}, C^{\mu,{\rm NP}}_{9'})$ scenario requires a breaking of GMFV~\footnote{In this contribution we focus on the scenarios compatible with GMFV, for a discussion regarding the $(C^{\mu,{\rm NP}}_{9}, C^{\mu,{\rm NP}}_{9'})$ scenarios see Ref.~\cite{Descotes-Genon:2020buf}}. Since the parameter combinations entering the charged and neutral lepton decays differ ($C_S+C_T$ vs $C_S-C_T$), a single point in the $\btosll$ parameter space defines a curve or a region in the \btosnunu and \stodnunu parameter space.  Further constraints can only be derived in specific scenarios allowing us to convert the information from $b \to s \ell^+ \ell^-$ observables into a constraint on $C_S$ and $C_T$ the simplest possibilities being $C_S=0$ or $C_T=0$.

\begin{figure}[h]
    \centering
    \includegraphics[width=0.365\textwidth]{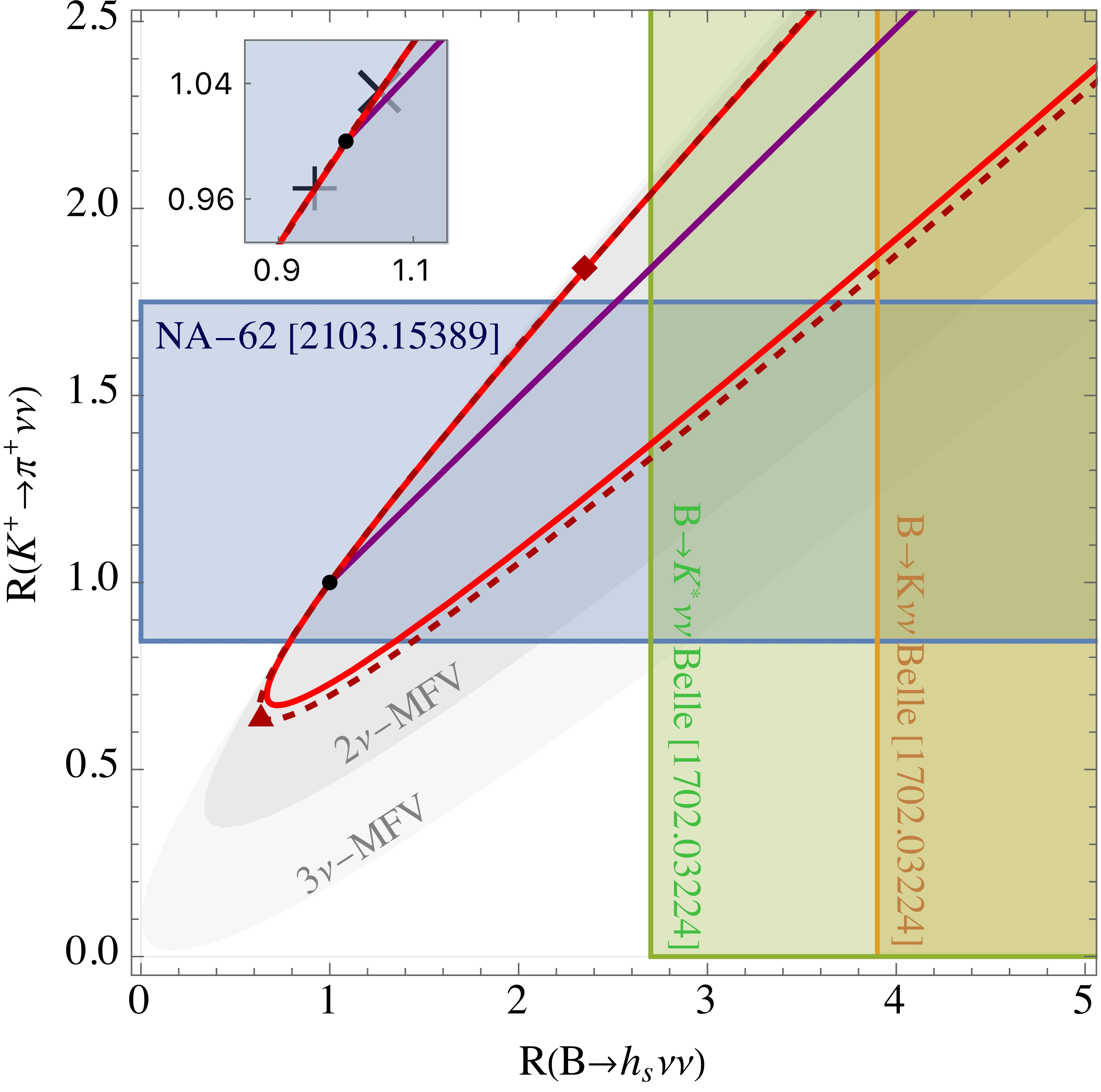}
    \hspace{3em}
      \includegraphics[width=0.43\textwidth]{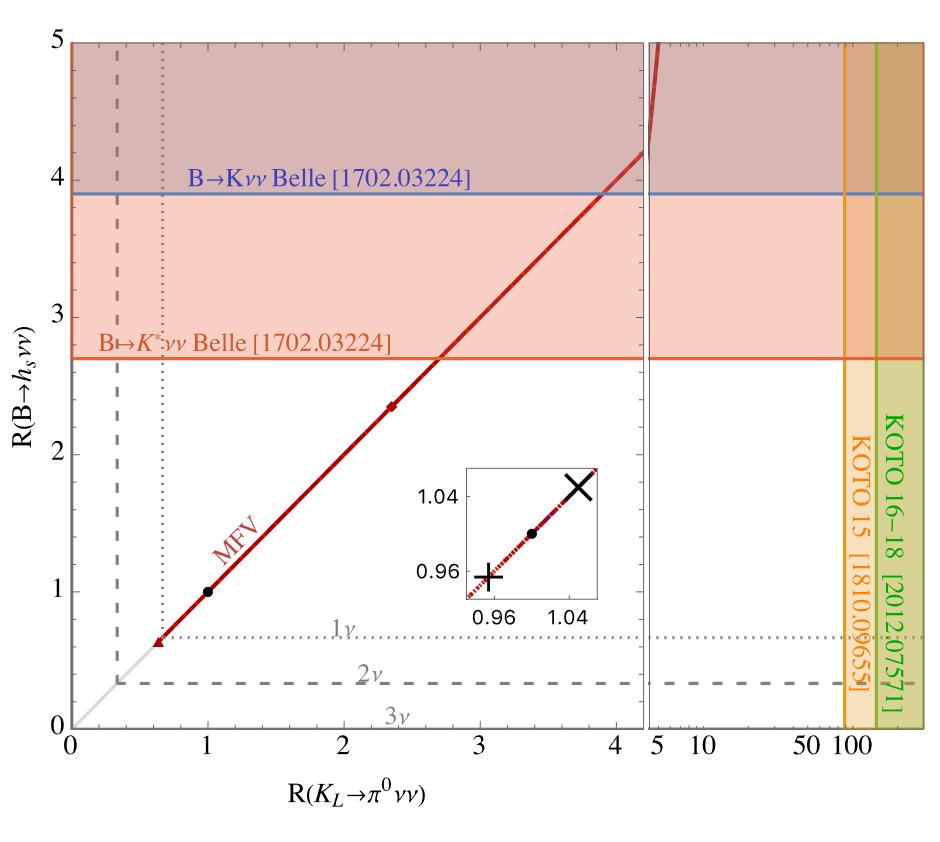}
    \caption{Correlation in the MFV limit between the ratios $R(B\to h_s\nu\bar\nu)$ ($h_s=K,K^*,X_s$) and $R(K^+\to \pi^+\nu\bar\nu)$ (left) and for the  ratios $R(B\to h_s\nu\bar\nu)$ ($h_s=K,K^*,X_s$) and $R(K_L\to \pi^0\nu\bar\nu)$ (right). }
    \label{fig:C9C10}
  \end{figure}

\section{Correlations between neutrino semileptonic modes} 

We first consider the limit of (linear) MFV in which $b\to s\nu\bar\nu$ and $s\to d\nu\bar\nu$ FCNC transitions are rigidly correlated through their dependence on $C_S-C_T$, even before considering the implications of $b\to s\ell^+\ell^-$.
In Fig.~\ref{fig:C9C10}, the left (right) figure shows the branching ratio of $B\to h_s \nu\nu$ and $K^+\to \pi^+\nu\bar\nu$ ($K_L\to \pi^0\nu\bar\nu$) normalised to their respectively SM value  $R(i\to f) \equiv \mathcal B(i\to f) /  \mathcal B(i\to f)_{\rm SM}$ and the constrains in the different NP scenarios. 

The allowed regions for the ratios of $B\to h_s \nu\nu$ and $K^+\to \pi^+\nu\bar\nu$ is shown shaded for arbitrary MFV NP effects on two (dark grey, 2$\nu$) or three (light grey, 3$\nu$) neutrino flavours. 
	The red, purple and dashed brown curves correspond to the allowed 1D regions for the three specific $U(1)^3_\ell$ scenarios Sc. 1, 2 and 3 respectively. The \btosll constraints  given $C_S=0$ or $C_T=0$  are indicated as black $\times$ and $+$ respectively in the inset plot for scenario 1 and brown $\Diamond$ or $\bigtriangleup$ respectively  for scenario 3 (already being probed by $\mathcal B(K^+\to \pi^+\nu\bar\nu)$ measurement and close to being probed by $B \to K^{(*)} \nu\bar\nu$ at the $B$-factories) while scenario 2 shows no significant deviations.
	In the case of $R(B\to h_s\nu\bar\nu)$ ($h_s=K,K^*,X_s$) vs $R(K_L\to \pi^0\nu\bar\nu)$, the parabolic region collapses into a single diagonal line due to stronger correlations between these two modes (see the addendum of Ref.~\cite{Descotes-Genon:2020buf}). 
    All ratios $R$ are bounded by the same minimal value $(1-N_\nu/3)$ where $N_\nu$ is the number of neutrino flavours affected by arbitrary NP.
	Also shown are the present experimental constraints coming from NA62~\cite{CortinaGil:2021nts}, $B$-factories~\cite{Grygier:2017tzo} and KOTO\cite{KOTO:2020prk,KOTO:2018dsc}.
An interesting observation is that a pair of future $B \to h_s \nu\bar\nu$ and $K^+\to\pi^+ \nu \bar\nu$ rate measurements outside of the (albeit large) grey region would be a clear indication of non-MFV NP.
In the case of $B\to h_s \nu\nu$ and $K_L\to \pi^0\nu\bar\nu$ already a deviation from the diagonal would indicate non-MFV. 

\begin{figure}[h]
    \centering
    \includegraphics[width=0.43\textwidth]{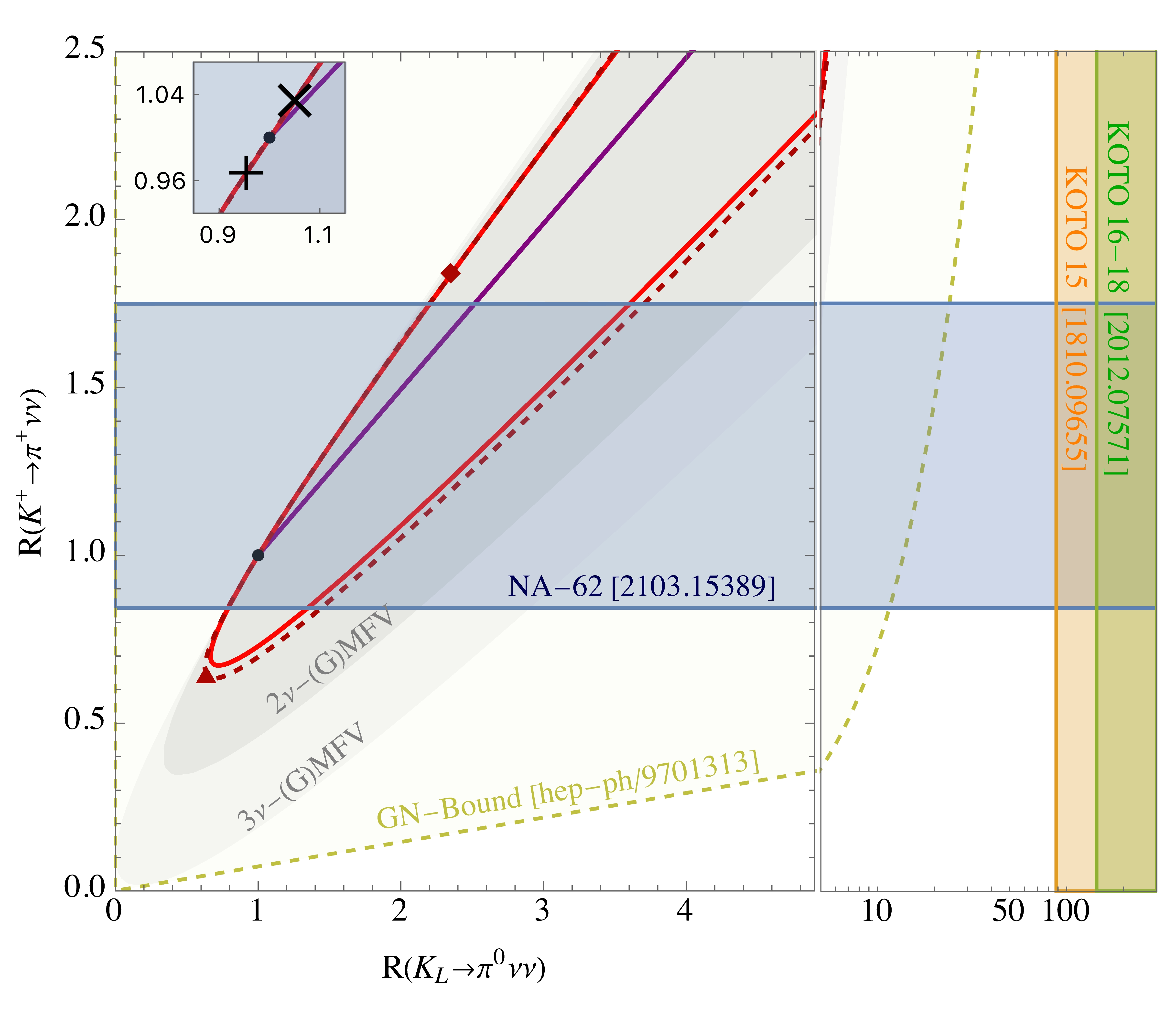}
    \caption{Correlation between the ratios $R(K^+\to \pi^+\nu\bar\nu)$  and $R(K_L\to \pi^0\nu\bar\nu)$ in the (G)MFV limit.
    Regions are the same as Fig.~\ref{fig:C9C10} except for the yellow region which corresponds to the Grossman-Nir Bound\cite{Grossman:1997sk}. }
    \label{fig:K+KL}
\end{figure}

In the more general case of GMFV a similar discussion can be done for the $K^+\to \pi^+\nu\bar\nu$ and $K_L\to \pi^0\nu\bar\nu$ modes. 
In Fig.~\ref{fig:K+KL}, an equivalent correlation to the one in Fig.~\ref{fig:C9C10} (left) can be seen for the GMFV case, relating both $\stodnunu$ modes. An additional constraint in Fig.~\ref{fig:K+KL} comes from the Grossman-Nir Bound\cite{Grossman:1997sk}.

  In summary, we have investigated, in a general EFT framework, the consequences of  $b\to s\mu\mu$ SM deviations in other FCNC processes. Under specific assumptions, we have studied the correlation between the branching ratios for $B\to h_s\nu\bar\nu$ , $K^+\to\pi^+\nu \bar\nu$ and  $K_L\to\pi^0\nu \bar\nu$ . The measurement of these modes could establish NP flavour breaking beyond (G)MFV as well as indicate the number of lepton flavours affected by NP.

\section*{Acknowledgments}
I would like to thank my collaborators S. Descotes-Genon, J. F. Kamenik, S. Fajfer with whom this work was developed.
This study has been carried out within the INFN project (Iniziativa Specifica) QFT-HEP.

\printbibliography

\end{document}